\documentclass[Blank]{webofc}
\usepackage[varg]{txfonts}   
\usepackage{float}
\usepackage{amsmath}
\usepackage{subfigure}
\usepackage{titlesec}
\titlespacing\section{0pt}{12pt plus 4pt minus 2pt}{5pt plus 2pt minus 2pt}
\titlespacing\subsection{0pt}{5pt plus 4pt minus 2pt}{5pt plus 2pt minus 2pt}
\titlespacing\subsubsection{0pt}{5pt plus 4pt minus 2pt}{5pt plus 2pt minus 2pt}

\newcommand\aap{A\&A}%
          
\woctitle{$6^{th}$ Roma International Conference on AstroParticle Physics}
\begin{document}
\title{Limits on Lorentz invariance violation at the Planck energy scale from H.E.S.S. spectral analysis of the blazar Mrk 501}
%
%

\author{\firstname{Matthias} \lastname{Lorentz}\inst{1}\fnsep\thanks{\email{matthias.lorentz@cea.fr}} \and
        \firstname{Pierre} \lastname{Brun}\inst{1}\fnsep\thanks{\email{pierre.brun@cea.fr}}
        , for the H.E.S.S. collaboration.
        }

\institute{ Irfu, CEA Saclay, 91191 Gif-sur-Yvette France}

\abstract{Some extensions to the Standard Model lead to the introduction of Lorentz symmetry breaking terms, expected to induce deviations from Lorentz symmetry around the Planck scale. A parameterization of effects due to Lorentz invariance violation (LIV) can be introduced by adding an effective term to the photon dispersion relation. This affects the kinematics of electron-positron pair creation by TeV $\gamma$ rays on the extragalactic background light (EBL) and translates into modifications of the standard EBL opacity for the TeV photon spectra of extragalactic sources. Exclusion limits are presented, obtained with the spectral analysis of H.E.S.S. observations taken on the blazar Mrk 501 during the exceptional 2014 flare. The energy spectrum, extending very significantly above 10 TeV, allows to place strong limits on LIV in the photon sector at the level of the Planck energy scale for linear perturbations in the photon dispersion relation, and provides the strongest constraints presently for the case of quadratic perturbations.}

\maketitle
\section{Introduction}
\label{intro}

Special relativity is a pillar of modern physics and Lorentz symmetry has been established to be an exact symmetry of Nature up to the precision of current experiments. It has been suggested however that this symmetry could only be approximate and that deviations from Lorentz invariance could appear at an energy scale beyond our current grasp. A generic approach to introduce such effects consists of adding effective terms in the dispersion relation of particles, \textit{i.e.} for photons

\begin{equation}
E_\gamma^2 = p_\gamma^2 \pm E_\gamma^2  \left( \frac{E_\gamma}{E_{\text{LIV}}} \right)^n,
\label{LIV_Dispersion}
\end{equation}
where $E_{\text{LIV}}$ is the hypothetical energy scale at which Lorentz symmetry could stop being exact, and $n$ the order of the leading correction. In some approaches to quantum gravity $E_{\text{LIV}}$ is expected to be of the order of Planck energy  $\text{E}_{\text{Planck}} = \sqrt{\hbar c^5 /G} \simeq 1.22 \times 10^{28} \text{ \ eV}$ \cite{QG}. As such deviations are only expected for photons at the highest energies, astrophysical $\gamma$-rays can be used to probe potential LIV effects. The most widely-used approach is to look for energy-dependent time delays for photons produced by distant $\gamma$-ray bursts (GRB) or during TeV flares of active galaxy nuclei (AGN) , see \textit{e.g.} \cite{HESS_TimeLags_PKS}. An attractive alternative possibility takes advantage of the fact that the modified dispersion relation for photons that could be induced by LIV would affect the kinematics for the $e^+e^-$ pair production of TeV $\gamma$ rays coming from AGNs on the EBL resulting in a modified opacity to extragalactic $\gamma$ rays, see \textit{e.g.} \cite{JacobPiran_2008}. In the following we consider LIV affecting only photons (like in \cite{Fairbairn_CTA} \cite{TavecchioBonnoli_2016}), not electrons as the constraints on LIV for electrons are very stringent due to observations of synchrotron radiation from the Crab Nebula \cite{Liberati_LIV_e}.

\section{Modified EBL opacity in the presence of LIV}
\begin{figure}[H]
\center
\includegraphics[scale=0.27]{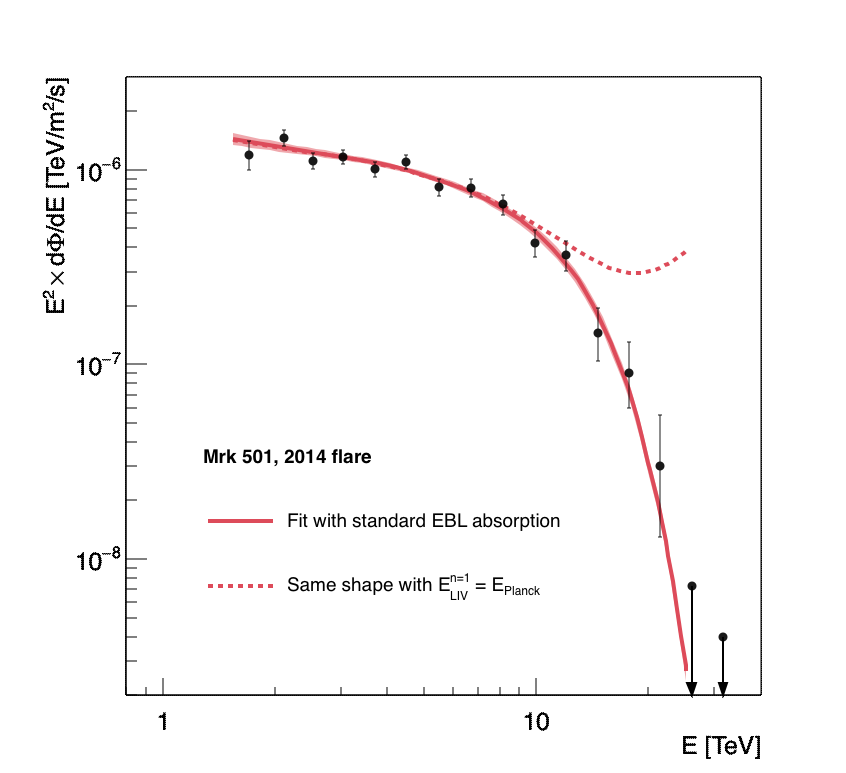}
\caption{Energy spectrum of Mrk 501 obtained from the H.E.S.S. phase-I analysis of the 2014 flare data. The fitted EBL-absorbed power law for the standard case is showed by the solid line, as well with the corresponding $1-\sigma$ confidence band. For comparison the same intrinsic power law with modified EBL absorption due to linear Planck scale perturbations is represented by the dashed line.}
\label{LIVSpec}
\end{figure}

The EBL is the background photon field originating from the integrated starlight and its re-processing by the interstellar medium over cosmic history. Its spectral energy distribution has two main components, an optical ($\sim 1$  eV) and an infrared ($\sim 10^{-2}$ eV) component. Extragalactic very high-energy (VHE, E > $100$ GeV) $\gamma$ rays can be used as an independent way to probe this background radiation, as such $\gamma$ rays interact with EBL photons via $e^+e^-$ pair production \cite{Gould67}, resulting in an attenuated observed flux (for a review see \cite{Dwek2013}).
The optical depth for a VHE photon of energy $E_\gamma$ traveling through a medium with EBL physical density $n( \epsilon, z)$ from a source at $z_s$ is:

\begin{equation}
\tau(E_\gamma,z_s) =c \int_0^{z_s} dz \frac{dt}{dz}  \int_0^2 d\mu \frac{\mu}{2} \int_{\epsilon_{thr}}^\infty d \epsilon \frac{dn(\epsilon,z)}{d\epsilon} \sigma_{\gamma \gamma} \left(E_\gamma(1+z), \epsilon, \mu \right),
\label{Tau}
\end{equation}

where $dt/dz=\left( H_0 (1+z) \sqrt{\Omega_{\text{M}}(1+z)^3 +\Omega_\Lambda}\right) ^{-1}$ \footnote{We assume a flat $\Lambda CDM$ cosmology with $\Omega_{\text{M}}=0.3$, $\Omega_\Lambda=0.7$ and $H_0=70 \text{ km } s^{-1} \text{ Mpc}^{-1}$. }, $\mu=1- \cos(\theta)$, $\epsilon_{thr} (E_\gamma,z)= \frac{2 m_e^2 c^4}{ E_\gamma \mu(1+z)}$ and $\sigma_{\gamma \gamma}$ is the Bethe-Heitler cross section for pair production. The absorption effect on the intrinsic spectrum of an extragalactic source is expressed as $\Phi_{\text{obs}}(E_\gamma)= \Phi_{\text{int}}(E_\gamma) \text{ e}^{-\tau(E_\gamma,z_s)}$. EBL absorption then leaves a typical redshift and energy-dependent imprint on the observed spectrum of extragalactic sources.
Knowledge of the EBL spectral energy distribution has greatly improved over the last decade, constraints from VHE $\gamma$ rays (see \textit{e.g.} \cite{Biteau12} \cite{Biteau15} \cite{Lorentz15}), predictions from models (see \textit{e.g.} \cite{Fr08} \cite{Dominguez11}), and results from an empirical determination \cite{Stecker:2016fsg} agree in between lower and upper limits.\\
The effective dispersion relation in the presence of LIV Eq.\ref{LIV_Dispersion} propagates into the optical depth given in Eq.\ref{Tau}, the invariant center-of-mass energy squared $s$ and threshold energy $\epsilon_{\text{thr}}$ become :

\begin{equation}
 s \rightarrow s \pm \frac{E_\gamma ^{n+2} }{E_{\text{LIV}}^n} \text{, and    } \epsilon_{\text{thr}} \rightarrow \epsilon_{\text{thr}} \mp \frac{1}{4} \frac{E_\gamma ^{n+1} }{E_{\text{LIV}}^n}
\end{equation}
\noindent
We assume, as in \cite{Fairbairn_CTA}, that the modified expression of $s$ can still be considered as an invariant quantity in the LIV framework (for a discussion see Appendix A. in \cite{TavecchioBonnoli_2016}). We only consider the subluminal case (minus sign in Eq. \ref{LIV_Dispersion}) : if non negligible, the effective term will induce lower values for $s$ suppressing pair creation on the EBL, causing an excess of transparency for $\gamma$ rays \footnote{Alternatively, in the superluminal case the threshold energy would be lowered implying an enhanced pair production for $\gamma$ rays that could more easily interact with the cosmic microwave background. This would result in a strong cut-off in the observed energy spectra of AGNs. This LIV scenario is unlikely with respect to current observations and also theoretically disfavored.}, see Fig. \ref{LIVSpec}.

\section{H.E.S.S. observations of Mrk 501 during the 2014 flare }
\subsection{H.E.S.S. experiment}
The High Energy Stereoscopic System (H.E.S.S.) is an array of five imaging atmospheric Cherenkov telescopes located 1800 m above see level in the Khomas Highland, Namibia, detecting $\gamma$-rays ranging from $\sim$ 100 GeV to a few tens of TeV. This is precisely the reciprocal sensitivity range for absorption due to the EBL for intermediate redshifts (z < 1).

\subsection{Mrk 501}
Mrk 501 is a well known AGN at redshift $z = 0.034$ which belongs to the class of blazars, \textit{i.e.} with its relativistic jet closely aligned to our line of sight. It is known to be highly variable from radio to VHE $\gamma$ rays and is referred to as a high-frequency-peaked blazar with a flux-dependent spectral hardening observed during flaring states. Its spectral characteristics and its relatively low redshift allow for the detection of among the most energetic extragalactic $\gamma$ rays, making this source ideal to investigate LIV through spectral studies as it has already been done \citep{Stecker:2001vb} \cite{Biteau15} \cite{TavecchioBonnoli_2016} with the historically highest VHE flux recorded in 1997 by the HEGRA \cite{HEGRA_97} and CAT \cite{CAT_97} telescopes.

\subsection{Flare data set}
The 2014 H.E.S.S. observations of Mrk 501 were triggered following high fluxes reported by the FACT collaboration. Observations taken during the night of June 23-24 2014 (MJD 56831-56832) revealed an exceptional flare with highest fluxes of Mrk 501 ever recorded by H.E.S.S. \cite{ICRC_Var}. These observations were performed with full array of all five telescopes, however for this study requiring optimal sensitivity at highest energies, data from the central large telescope are not essential. The mean zenith angle of observations was $\sim 63 ^\circ$. The Model analysis  with loose cuts  \cite{ModelAnalysis} was performed leading to an excess of more than 1200 photons with a $\sim 67 \sigma$ significance for the 2 hours of observations taken that night.
Spectral analysis was performed using the forward folding technique described in \cite{FFolding_Piron}. The spectrum, extending significantly up to $\sim 20$ TeV, is well fitted ($\chi^2/\text{n.d.f} = 8.5/8$) by a simple EBL-absorbed power law using the EBL model of \cite{Fr08}. There is no evidence for intrinsic curvature nor cut-off. The fitted intrinsic spectral parameters read
\begin{equation}
\frac{d\Phi_{\text{int}}}{dE} = (1.68 \pm 0.16) \times 10^{-6} \left( \frac{E}{1 \text{ TeV}} \right)^{-2.15 \pm 0.06} \text{ m}^{-2} \text{ s}^{-1} \text{ TeV}^{-1}.
\end{equation}

\section{Results and discussion}
The maximum likelihood forward folding method for spectrum determination is performed assuming an intrinsic power law absorbed with the EBL model of \cite{Fr08}. The optical depths are computed considering modifications due to LIV as explained in Sec. 2. Values of $E_{\text{LIV}}$ are scanned logarithmically in the range
of interest for linear (n=1) and quadratic (n=2) scenarios. Log-likelihood profiles for both cases are 
shown in Fig. 2. As the data show no evidence for a high-energy upturn, the fit prefers LIV-free optical
depth values. Indeed log-likelihood values reach plateaus corresponding to the standard case with no deviations from Lorentz symmetry in both cases. This allows to compute exclusion limits on $E_{\text{LIV}}$, as summarized in Tab. \ref{LimitsTable}.  

\begin{figure}[H]
\hfill
\subfigure[Linear case]{\includegraphics[scale=0.23]{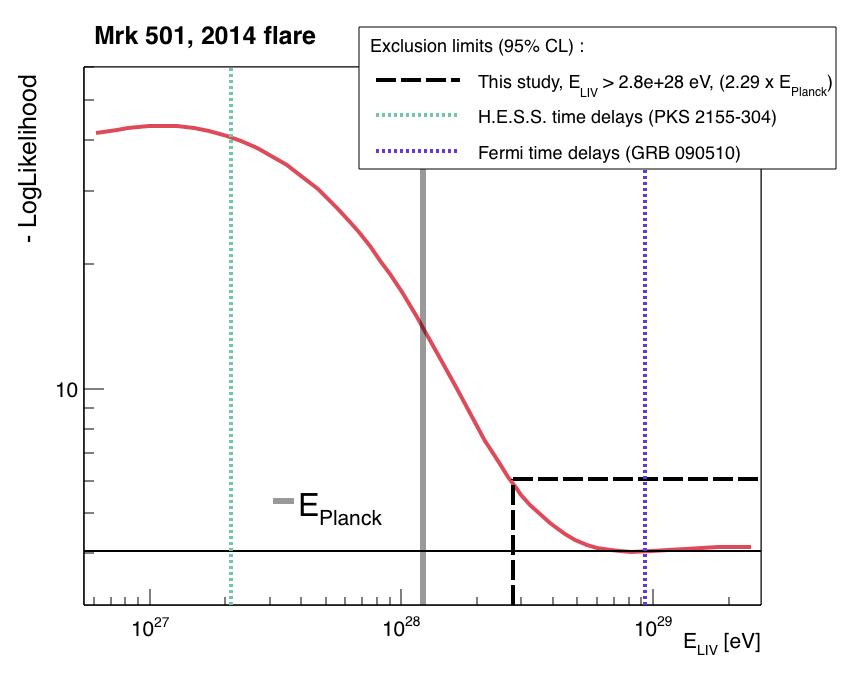}}
\hfill
\subfigure[Quadratic case]{\includegraphics[scale=0.23]{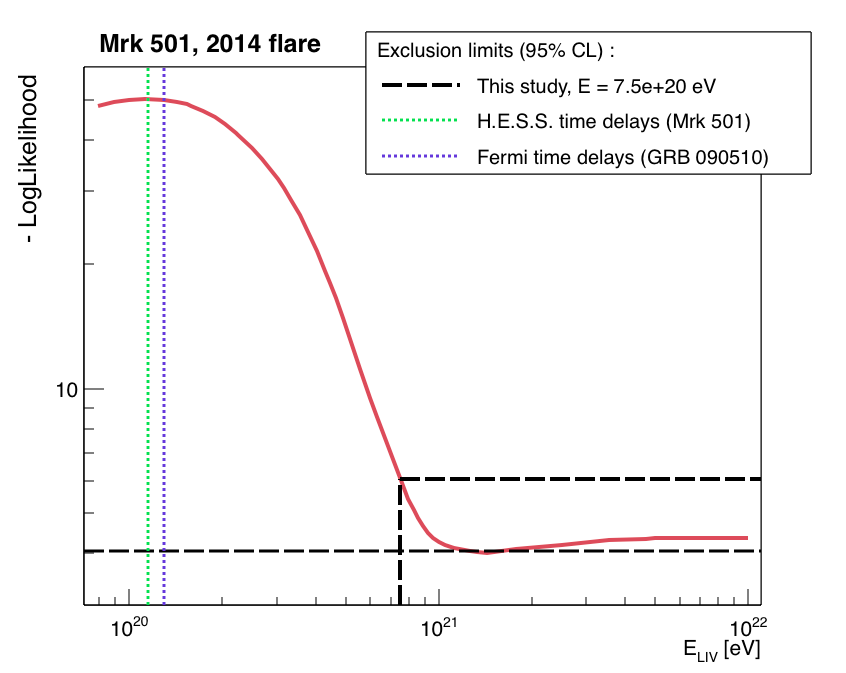}}
\hfill
\caption{Log-likelihood profiles and corresponding 95\% confidence level (CL) exclusion limits for the $E_{\text{LIV}}$ scan in the linear (left) and quadratic (right) case. Also showed are the best limits obtained with the method of energy-dependent time delays with AGNs \cite{HESS_TimeLags_PKS} and GRBs \cite{Fermi_TimeLags_GRB}. }
\label{LProfiles}
\end{figure}

{\renewcommand{\arraystretch}{1.2}
\begin{table}[H]
\centering
\begin{tabular}{llllll}
\cline{1-4}
\multicolumn{1}{|l|}{}    & \multicolumn{1}{c|}{$2 \ \sigma $} & \multicolumn{1}{c|}{$3 \ \sigma$} & \multicolumn{1}{c|}{$5 \ \sigma$} &  &  \\ \cline{1-4}
\multicolumn{1}{|l|}{n=1} & \multicolumn{1}{l|}{$2.8 \times 10^{28}$ eV ($2.29 \times \text{E}_{\text{Planck}}$)}       & \multicolumn{1}{l|}{$1.9 \times 10^{28}$ eV ($1.6 \times \text{E}_{\text{Planck}}$)} & \multicolumn{1}{l|}{$1.04 \times 10^{28}$ eV ($0.86 \times \text{E}_{\text{Planck}}$)}        &  &  \\ \cline{1-4}
\multicolumn{1}{|l|}{n=2} & \multicolumn{1}{l|}{$7.5 \times 10^{20}$ eV }        & \multicolumn{1}{l|}{$6.4 \times 10^{20}$ eV}   & \multicolumn{1}{l|}{$4.7 \times 10^{20}$ eV }      &  &  \\ \cline{1-4}
                          &                              &                      &        &  & 
\end{tabular}
\caption{Exclusion limits on $E_{\text{LIV}}$ obtained from the profiles of Fig. \ref{LProfiles}}
\label{LimitsTable}
\end{table}
}

For the linear case, the 95 $\%$ CL limit is at $2.8 \times 10^{28}$ eV (at $\sim$ $2.3\times \text{E}_{\text{Planck}}$), one order of magnitude above the current best limit using timelags for AGNs \cite{HESS_TimeLags_PKS}, and below the best limit obtained with GRBs \cite{Fermi_TimeLags_GRB}.
The 5-$\sigma$ exclusion is at $1.044\times 10^{28}$ eV ($0.86 \times \text{E}_{\text{Planck}}$) and $\text{E}_{\text{Planck}}$ is excluded at the 4.5 $\sigma$ level. For the quadratic case the 95 $\%$ CL limit is at $7.5 \times 10^{20}$ eV, more than 6 times above current best timelag limits with AGNs and GRBs. The 5-$\sigma$ exclusion is at $4.7 \times 10^{20}$ eV. These are the best current exclusion limits in the quadratic case.\\
These strong constraints naturally come from the exceptional spectrum of the 2014 flare data-set where the power law intrinsic emission extends up to 20 TeV.

A cross-check analysis using an independent simulation of the shower development in the atmosphere and independent calibration and reconstruction chains lead to a compatible spectrum. This gives good confidence that this spectrum, on which the LIV analysis is built, is not affected by large systematic uncertainties. The same LIV analysis using the EBL model of \cite{Dominguez11} leads to very similar exclusion limits, showing that the choice of a specific EBL model does not significantly affect our results.

\section{Conclusions}
The non observation of deviations from standard EBL absorption in the multi-TeV spectrum of Mrk 501 observed by H.E.S.S. during the 2014 flare allows us to derive strong limits on $E_{\text{LIV}}$ in the photon sector, currently the best limits obtained with an AGN.
This confirms the result obtained with GRB 090510 that standard photon dispersion relation holds up to the Planck energy scale in the case of linear perturbations, and pushes higher the current limit in the case of quadratic perturbations.

\section*{Acknowledgments}
\noindent
Please see standard acknowledgements in H.E.S.S. papers, not reproduced here due to lack of space.
%
%
%

\end{document}